\documentclass[usegraphicx,usenatbib,useAMS]{mn2e}
\usepackage{amsmath}
\usepackage{amssymb}
\usepackage{color}
\usepackage{threeparttable}

\newcommand{\lya}{\ifmmode\mathrm{Ly}\alpha\else{}Ly$\alpha$\fi}
\newcommand{\lyb}{\ifmmode\mathrm{Ly}\beta\else{}Ly$\beta$\fi}
\newcommand{\igm}{\ifmmode\mathrm{IGM}\else{}IGM\fi}
\newcommand{\lae}{\ifmmode\mathrm{LAE}\else{}LAE\fi}
\newcommand{\bao}{\ifmmode\mathrm{BAO}\else{}BAO\fi}
\newcommand{\h}{\ifmmode\mathrm{H}\else{}H\fi}
\newcommand{\hi}{\ifmmode\mathrm{H\,{\scriptscriptstyle I}}\else{}H\,{\scriptsize I}\fi}
\newcommand{\hii}{\ifmmode\mathrm{H\,{\scriptscriptstyle II}}\else{}H\,{\scriptsize II}\fi}
\newcommand{\he}{\ifmmode\mathrm{He}\else{}He\fi}
\newcommand{\hei}{\ifmmode\mathrm{He\,{\scriptscriptstyle I}}\else{}He\,{\scriptsize I}\fi}
\newcommand{\heii}{\ifmmode\mathrm{He\,{\scriptscriptstyle II}}\else{}He\,{\scriptsize II}\fi}
\newcommand{\heiii}{\ifmmode\mathrm{He\,{\scriptscriptstyle III}}\else{}He\,{\scriptsize III}\fi}
\newcommand{\cmb}{\ifmmode\mathrm{CMB}\else{}CMB\fi}
\newcommand{\qso}{\ifmmode\mathrm{QSO}\else{}QSO\fi}
\newcommand{\civ}{\ifmmode\mathrm{C\,{\scriptscriptstyle IV}}\else{}C\,{\scriptsize IV}\fi}
\newcommand{\siv}{\ifmmode\mathrm{Si\,{\scriptscriptstyle IV}}\else{}Si\,{\scriptsize IV}\fi}
\newcommand{\ovi}{\ifmmode\mathrm{O\,{\scriptscriptstyle VI}}\else{}O\,{\scriptsize VI}\fi}

\title[Temperature fluctuations in the \lya{} forest]{The impact of
  temperature fluctuations on the large-scale clustering of the
  \lya{} forest}

\author[B. Greig, J.S. Bolton and J.S.B. Wyithe] {Bradley~Greig,$^{1,2,3}$\thanks{E-mail:~b.greig@student.unimelb.edu.au~(BG),}, James S.~Bolton$^{3, 4}$ and J. Stuart B.~Wyithe$^{2,3}$ \\ 
$^1$Scuola Normale Superiore, Piazza dei Cavalieri 7, 56126 Pisa, Italy \\
$^2$School of Physics, University of Melbourne, Parkville, Victoria 3010, Australia\\
$^3$ARC Centre of Excellence for All-sky Astrophysics (CAASTRO), 44-70 Rosehill Street, Redfern NSW 2016, Sydney, Australia\\
$^4$School of Physics and Astronomy, University of Nottingham, University Park, Nottingham, NG7 2RD, UK}

\begin{document}
\maketitle \begin{abstract} We develop a semi-analytic method for
  assessing the impact of the large-scale \igm{} temperature fluctuations
  expected following \heii{} reionization on three-dimensional
  clustering measurements of the \lya{} forest.  Our methodology
  builds upon the existing large volume, mock \lya{} forest survey
  simulations presented by Greig et al.\ by including a
  prescription for a spatially inhomogeneous ionizing background,
  temperature fluctuations induced by patchy \heii{} photoheating and
  the clustering of quasars. This approach enables us to achieve
  a dynamic range within our semi-analytic model substantially larger 
  than currently feasible with computationally expensive, fully numerical 
  simulations. The results agree well with existing
  numerical simulations, with large-scale temperature fluctuations
  introducing a scale-dependent increase in the spherically averaged
  3D \lya{} forest power spectrum of up to 20$-$30 per cent at
  wavenumbers $k\sim 0.02\rm\,Mpc^{-1}$.  Although these large-scale
  thermal fluctuations will not substantially impact upon the
  recovery of the baryon acoustic oscillation scale from existing and
  forthcoming dark energy spectroscopic surveys, any complete forward
  modelling of the broad-band term in the \lya{} correlation function
  will none the less require their inclusion.

\noindent

\end{abstract} 

\section{Introduction}

The \lya{} forest is a powerful probe of large-scale structure in the
intergalactic medium (\igm{}).  It is characterized by a series of
narrow absorption features imprinted into the spectra of bright
background sources, and arises from the resonant scattering of \lya{}
photons by intervening neutral hydrogen \citep{Rauch:1998p4563}.  The
observed transmitted flux yields information on the underlying
physical properties of the \igm{}, such as the gas density, ionization
state and temperature, and is sensitive to the matter power spectrum
at small scales, $1$--$80\,h^{-1}\rm\,Mpc$, along the line-of-sight
\citep{Croft:2002p6751,Viel:2004p7045,McDonald:2006p6764,Seljak:2006p14,Palanque-Delabrouille2013}.

Recently, \citet[][see also \citealt{Slosar:2013p15454,Delubac:2014p1801,Ribera:2014p27}]{Busca:2012p13850}
reported the first successful detection of baryon acoustic oscillations
(\bao{}s) in the \lya{} forest from the Baryon Oscillation
Spectroscopic Survey (BOSS).  These observations have highlighted the
potential of the \lya{} forest for probing the \emph{three-dimensional} 
large-scale clustering of structure within the \igm{},
in addition to the existing 1D line-of-sight measurements.  However,
these data (as well as results from forthcoming surveys such as eBOSS
and DESI) will also be sensitive to large-scale fluctuations of
\emph{astrophysical} origin.  Large-scale variations in the ionization
and thermal state of the \igm{} in particular may be detectable at
$2<z<3$ \citep{Slosar:2009p5710,White:2010p5647,McQuinn:2011p10618}.
Indeed, both \citet{Pontzen:2014p0506} and \citet{Gontcho:2014p7425}
have recently developed analytical models for investigating the impact
of very large-scale ($k<0.01\,h\,{\rm Mpc}^{-1}$) ionization fluctuations
on the \lya{} forest.  At the mean BOSS redshift of $z\sim2.3$, these
studies concluded there should be a measurable impact on the \hi{}
power spectrum at large scales as a result of the interplay between
the clustering of the \igm{} and fluctuations in the underlying \hi{}
fraction.

However, significant fluctuations in the \igm{} temperature of the order of
$\sim 10^{4}\rm\,K$ are also expected to linger following the
reionization of singly ionized helium (\heii{}) at $z\sim 3$
\citep{McQuinn:2009p7230,Compostella:2013p5745}. These will result in
large-scale variations in the \hi{} fraction through the temperature
dependence of the \hii{} recombination coefficient, $\alpha_{\rm
  HII}\propto T^{-0.7}$.  \citet{Gontcho:2014p7425} briefly discussed
the impact that these temperature fluctuations might have on the
\lya{} power spectrum, although as noted by these authors, their
analytic model was not suitable for fully modelling both the local and
global effects of heating during \heii{} reionization.  Earlier work
by \citet{McQuinn:2011p10618} also investigated the signature of
temperature fluctuations in the \igm{} using an analytical argument
combined with radiative transfer simulations (RT) of \heii{} reionization
\citep{McQuinn:2009p7230}.  These authors found order of unity
variations in the amplitude of the 3D \lya{} forest power spectrum at
$k<0.1\rm\,Mpc^{-1}$.  However, the simulations used to perform the
\citet{McQuinn:2011p10618} analysis were not large enough to sample
fluctuations on scales $\la 0.05 \rm\,Mpc^{-1}$, nor were the \lya{}
forest spectra extracted from the RT simulations able
to resolve the Jeans scale at mean density.  Self-consistently
modelling the expected signature of temperature fluctuations in \lya{}
spectroscopic surveys is thus extremely computationally challenging.
A large dynamic range is required to simultaneously resolve
small-scale structure in the \lya{} forest and probe sufficiently
large cosmic volumes which incorporate the sparse ionizing sources
(quasars; e.g. \citealt{Furlanetto:2008p4660}) thought to drive the
temperature fluctuations.

An alternative approach is therefore to use simplified but
significantly more efficient techniques to investigate this
problem. In this context, \citet[][hereafter GBW11]{Greig:2011p13768}
developed an efficient, large-volume mock \lya{} forest survey model
which was used to simulate the signature of \bao{}s in the \lya{}
forest.  The results agreed well with a variety of observational
measurements after calibration against hydrodynamical simulations, and
were additionally able to sample large cosmological volumes ($\sim
1\rm\, Gpc^{3}$) at high resolution ($\sim 240\rm\, kpc$).  In this
work, we now extend this model to examine the effect of temperature
fluctuations following \heii{} reionization on measurements of
large-scale clustering with the \lya{} forest.  

The paper is organized as follows.  In Section~\ref{sec:Reion_Model},
we outline modifications made to our previous simulations and then
describe our model for large-scale temperature fluctuations in
Section~\ref{sec:IGMTemp}.  We then investigate the impact of these
large-scale thermal fluctuations on the \lya{} forest in
Section~\ref{sec:Analysis}.  Finally, in Section~\ref{sec:conclusion},
we summarize our results and provide our closing remarks.
Throughout this work we assume a $\Lambda$ cold dark matter
cosmology with, $h=0.72$, $\Omega_{\rm{m}} = 0.26$,
$\Omega_{\Lambda}=0.74$, $\Omega_{\rm{b}}=0.0444$, $n_{\rm{s}}=0.96$
and $\sigma_{8}=0.8$. These are equivalent to our calibration
hydrodynamical simulation, model L3 of \citet{Bolton:2010p6750}.

\section{Large-scale \lya{} forest simulations} \label{sec:Reion_Model}

Hydrodynamical simulations of the \igm{} have been enormously
  successful at reproducing the basic observational properties of the
  \lya{} forest (e.g. \citealt{Croft:1998p44,Theuns:1998p7245}).
  However, these simulations are often limited to relatively small
  volumes ($<40 \rm\,Mpc^{3}$) to ensure the spatial resolution is
  sufficient for fully resolving the Jeans scale in the low-density
  IGM.  Even state-of-the-art IGM simulations (e.g.\
  \citealt{Lukic2014}) still do not approach the very large (Gpc)
  scales required to study \bao{}s in the \lya{} forest. In the last
  few years, this has led to renewed interest in the development of
  simplified techniques for modelling large-scale correlations in the
  \lya{} forest.  These approaches achieve the required dynamic range,
  but often at the expense of accuracy on small scales where the
  density field is mildly non-linear.  For example,
  \cite{FontRibera:2012p1} generate \lya{} forest spectra by
  inexpensively generating one dimensional Gaussian fields and
  subsequently introducing three dimensional correlations into the
  mock data.  \cite{LeGoff:2011p1737} generate three dimensional
  Gaussian density fields and use a lognormal mapping to obtain the
  \igm{} density field.  Most recently, \cite{Peirani:2014p1703}
  utilized large-scale dark matter simulations and used conditional
  probability distributions to directly relate the \lya{} transmission
  to the dark matter density.  This latter method is more
  computationally demanding than the other approaches (although still
  less so than hydrodynamical simulations), but it has the advantage
  of capturing small-scale correlations more accurately.

The \lya{} forest models in this work are based on the fast, large
volume simulations developed by GBW11.  This is itself based on a
model developed by \citet{Viel:2002p4288}, whereby a linear density
field undergoes a rank-ordered mapping of its probability distribution
to the corresponding distribution from a calibration hydrodynamical
simulation. In this regard, our approach is most similar to
  that used by \cite{LeGoff:2011p1737}.  Although this model does not
correctly capture the mildly non-linear structure probed by the \lya{}
forest on small scales, it is very well suited for studying the
fluctuations on large scales considered here.  We direct the reader to
GBW11 for further discussion and tests of our \lya{} forest model.
Here, we restrict ourselves to describing only the new components we
have implemented for examining the effect of temperature fluctuations
on \lya{} forest clustering following \heii{} reionization.  Note that
all quantitative results in this work are obtained at redshift $z=2.5$
using $1\rm\,Gpc^{3}$ comoving simulation volumes with $4096^{3}$
pixels.

\subsection{Clustering of ionizing sources} \label{sec:QSOclust}

Quasars, with their hard non-thermal spectra, are the sources
considered most likely to drive the process of \heii{} reionization
\citep{Madau:1999p13717,Furlanetto:2008p4660,McQuinn:2009p7230}.
These sources are known to cluster on scales of a few tens of Mpc
\citep{Shen:2007p2222} and to be strongly biased tracers of the
large-scale structure.  Therefore, to develop a physically motivated
model for temperature fluctuations in the \igm{} following \heii{}
reionization, we must first identify the location of the quasars in
our simulation volume.

We achieve this by performing an iterative procedure to identify
suitable locations for the ionizing sources.  This begins with a large
($1\rm\,Gpc^{3}$) volume box which contains a gridded $4096^{3}$
linear \igm{} density field in Fourier space.  This is then smoothed
on a mass-scale of 5$\times10^{12}\,$M$_{\sun}$ by a top-hat filter in
real-space, roughly corresponding to the expected characteristic halo
mass hosting a luminous quasar.  After Fourier transforming the
density field to real space, a pixel is then flagged as a candidate
quasar location when (i) its smoothed linear density exceeds four
times the linear critical density for collapse and (ii) it is the
highest density peak within a grid centred on the candidate pixel
which extends 8 pixels from centre to edge.  Finally, we sort all
candidate quasar locations by their smoothed linear overdensity,
sampling from highest to lowest.  As a test of this procedure, we have
computed the quasar correlation function recovered by this approach
and verified there is excellent agreement to the observed clustering of
luminous quasars presented by \citet{Shen:2007p2222}.

\subsection{Spatial fluctuations in the ionizing background} \label{sec:UVBModel}

With our prescription for identifying the locations of the ionizing
sources in hand, we now turn to modelling the spatial fluctuations in
the ionizing background. We use the method outlined by
\citet{Bolton:2011p13588} for this purpose (see also
\citealt{Furlanetto:2009p13603}).  

First, we must assign luminosities to the quasar locations identified
in our simulation volume using the observed quasar luminosity function
(QLF).  The number of quasars to be placed within our simulation
volume, $V$, is obtained by integrating the $B$-band QLF,
$\phi(L_{B},z)$ of \citet{Hopkins:2007p13715},
\begin{eqnarray} \label{eq:QLF}
N(L_{B} > L_{\rm min}) = V\int^{\infty}_{L_{\rm min}}\phi(L_{B},z) {\rm d}L_{B},
\end{eqnarray}
where we assume a limiting magnitude of $L_{\rm min} = 10^{44.25}$ erg
s$^{-1}$ ($M_{B} = $\,\,--22) in this work.

We next assign each quasar a $B$-band luminosity by Monte Carlo
sampling the observed QLF, selecting from our sorted list of candidate
quasars until we obtain $N(L_{B} > L_{\rm min})$ sources.  We assume a
broken power-law spectral energy distribution
\citep{Madau:1999p13717},
\begin{eqnarray}
L_{\nu} = \left\{ \begin{array}{cl} 
      \nu^{-0.3}  &  ( 2500 < \lambda < 4600 {\rm\AA} ),\\ 
      \nu^{-0.8} &  ( 1050 < \lambda < 2500 {\rm\AA} ), \\
      \nu^{-\alpha_{s}} &  (\lambda < 1050 {\rm\AA} ),
\end{array} \right.
\end{eqnarray}
where $\alpha_{\rm s}$ is the extreme UV 
index. \citet{Telfer:2002p14077} obtained $\alpha_{\rm s} = 1.57
\pm 0.17$ from a sample of low-redshift, radio-quiet quasars and
additionally observed significant source-to-source variations,
characterized as a Gaussian distribution with a mean $\alpha_{\rm s}
\approx 1.6$ and a standard deviation of $0.86$. More recently,
\citet{Shull:2012p162} and \citet{Stevans2014} recovered a mean
spectral index of $\alpha_{\rm s} = 1.41$ from \emph{HST}-COS observations
at $z<1.5$.  We therefore encompass the large observed variability in
$\alpha_{\rm s}$ by Monte Carlo sampling a Gaussian with a mean of
$1.5$ and standard deviation $0.5$ over the range of three standard
deviations.

We next calculate $J(\bmath{x},\nu)\, [{\rm
    erg\,\,s^{-1}\,cm^{-2}\,Hz^{-1}\,sr^{-1}}]$, the specific
intensity of the ionizing background.  For the $N$ quasars in our
simulation volume, at frequencies between the \hi{} and \heii{}
photoionization edges, $\nu_{\hi{}} < \nu < \nu_{\heii{}}$, the total
contribution to the ionizing background is given by,
\begin{eqnarray} \label{eq:JHI}
J(\boldsymbol{x},\nu) = \frac{1}{4\upi}\sum^{N}_{i=1}\frac{L_{i}(\boldsymbol{x}_{i},\nu)}{4\upi|\boldsymbol{x}-\boldsymbol{x}_{i}|^{2}},
\end{eqnarray}
where $|\boldsymbol{x}-\boldsymbol{x}_{i}|$ is the distance at a point
$\boldsymbol{x}$ to quasar $i$.  We have assumed the \igm{} is optically
thin below the \heii{} ionization threshold; recently reported values
for the mean free path for \hi{} ionizing photons at $z=2.4$ are in
the range $\sim 500$--$860$ comoving Mpc
\citep{OMeara:2013p137,Rudie2013}, which is an appreciable fraction of
our $1\rm\,Gpc^{3}$ simulation volume. As a result, our
simulations do not capture the very large scale ($k<0.01\rm\,
Mpc^{-1}$) ionization fluctuations discussed recently by
\citet{Pontzen:2014p0506} and \citet{Gontcho:2014p7425}.

At frequencies above the \heii{} ionization threshold, $ \nu >
\nu_{\heii{}}$, the ionizing background is instead,
\begin{eqnarray} \label{eq:JHeII}
J(\boldsymbol{x},\nu) = \frac{1}{4\upi}\sum^{N}_{i=1}\frac{L_{i}(\boldsymbol{x}_{i},\nu)}{4\upi|\boldsymbol{x}-\boldsymbol{x}_{i}|^{2}} {\rm e}^{-\frac{|\boldsymbol{x}-\boldsymbol{x}_{i}|}{\lambda_{\heii{}}}\left(\frac{\nu}{\nu_{\heii{}}}\right)^{-3(\beta-1)}},
\end{eqnarray}
where $\lambda_{\heii}$ is the mean free path of ionizing photons at the
\heii{} ionization threshold, which is a highly uncertain quantity
that will vary substantially during \heii{} reionization.  Models
derived from the observed \hi{} column density distribution report
values in the range $\sim 90$--$114$ Mpc at $z=2.5$ to $\sim 15$--$36$
Mpc (comoving) at $z=3$ \citep{Khaire2013,Davies:2012p14012}.  In this
work, we assume $\lambda_{\heii} = 45$ comoving Mpc, which is similar
to the expected mean separation of luminous quasars at $z\sim 2.5$
\citep{Davies:2012p14012}.  This choice maximises the impact of
temperature fluctuations on our results at $z=2.5$; larger mean free
paths will result in smaller temperature fluctuations in our model.
Note, however, that since the adiabatic cooling time-scale in the 
low-density \igm{} is long, $t\sim H^{-1}$, thermal fluctuations imprinted
during \heii{} reionization at $z\sim 3$ -- when the mean free path
may be rather short -- will also persist to lower redshift, and so the
assumption of a smaller mean free path is likely reasonable (see also
\citealt{Gontcho:2014p7425}). We further approximate the column density
distribution for \heii{} absorbers in the diffuse \igm{} to follow a power
law with $\beta=1.5$.  While this will broadly capture the effect of
Poisson distributed absorbers on the frequency dependence of the mean
free path, the distribution will start to deviate from a single
power law when \hi{} and \heii{} become optically thick
\citep{Fardal:1998p2206,HaardtMadau2012}.

Finally, to calculate the specific intensity we evaluate
equations~(\ref{eq:JHI}) and~(\ref{eq:JHeII}) using 30 frequency bins
spanning the range 1--100\,Ryd.  The photoionization rates
[s$^{-1}$] are then
\begin{eqnarray} \label{eq:photoionise}
\Gamma_{i}(\boldsymbol{x}) = \int_{\nu_{i}}^{\infty}4\upi \sigma_{i}(\nu)J(\boldsymbol{x},\nu) \frac{{\rm d}\nu}{h_{\rm p}\nu},
\end{eqnarray}
and the photoheating rates [erg s$^{-1}$] are
\begin{eqnarray}
g_{i}(\boldsymbol{x}) = \int_{\nu_{i}}^{\infty}4\upi \sigma_{i}(\nu)h_{\rm p}(\nu-\nu_{i})J(\boldsymbol{x},\nu) \frac{{\rm d}\nu}{h_{\rm p}\nu}.
\end{eqnarray}
Here, the subscript $i$ denotes three species of hydrogen and helium
(\hi{}, \hei{} and \heii{}), $h_{\rm p}$ is the Planck constant,
$\sigma_{i}(\nu)$ is the photoionization cross-section and $\nu_{\rm
  i}$ is the ionization threshold.

Assuming neutral hydrogen is in photo-ionization equilibrium, the
\hi{} number density at any location within our simulation volume is
then obtained via,
\begin{eqnarray}
n_{\hi{}}(\boldsymbol{x},z) & = & \frac{\alpha_{\hii{}}(T)}{\Gamma_{\hi{}}(\boldsymbol{x})}\frac{2-Y}{2(1-Y)}n^{2}_{\h}(\boldsymbol{x},z). \label{eq:nHI_new}
\end{eqnarray}
Here, $\alpha_{\hii{}}(T)$ is the temperature dependent recombination
rate, $\Gamma_{\hi{}}(\boldsymbol{x})$ is the spatially varying
photoionization rate determined from equation~(\ref{eq:photoionise})
and $Y$ is the helium mass fraction, for which we assume $Y = 0.24$
\citep{PlanckCollaboration:2013p15401}.

\subsection{Patchy heating from \heii{} reionization} \label{sec:Tboost}

For the final part of our semi-analytical patchy heating model, we use
the above calculation of the inhomogeneous ionizing background at
$E>4\rm\,Ryd$ to estimate the temperature of the \igm{} following \heii{}
reionization (see also \citealt{Raskutti2012,LidzMalloy2014} for
related approximate approaches for heating during \hi{} reionization
at $z>5$).

During \heii{} reionization, the excess energy of an \heii{} ionizing
photon above the ionization threshold goes into heating the \igm{},
resulting in a substantial boost to the overall temperature. The
precise amount of heating will depend on the intrinsic spectral shape
of the ionizing background and the degree of hardening as the
radiation is filtered through the \igm{}.  In practice, the excess energy
per photoionization for a power-law spectrum, $L_{\nu} \propto \nu^{-\alpha_{\rm s}}$,
will vary between the optically thin and thick
limits, $\langle E \rangle _{\rm thin}=h_{\rm p}\nu_{\rm
  HeII}/(\alpha_{\rm s}+2)$ to $\langle E\rangle _{\rm thick}=h_{\rm
  p}\nu_{\rm HeII}/(\alpha_{\rm s}-1)$ \citep{Abel:1999pL13}.
  
We may thus estimate the expected temperature boost in the \igm{}
following \heii{} reionization as \citep{Furlanetto:2008p13605}
\begin{eqnarray} \label{eq:Tboost}
\Delta T(\boldsymbol{x}) = 0.035f_{\heii{}}\left(\frac{2}{3k_{{\rm B}}}\right) \langle E_{\heii{}}(\boldsymbol{x}\rangle).
\end{eqnarray}
Here, $\langle E_{\heii{}}(\boldsymbol{x}) \rangle =
g_{\heii{}}(\boldsymbol{x})/\Gamma_{\heii{}}(\boldsymbol{x})$ is the
average excess energy per \heii{} photoionization (which will vary
between $\langle E_{\rm thin}\rangle \sim 15.5 \rm\,eV$ and $\langle
E_{\rm thick}\rangle \sim 108.8\rm \,eV$ for $\alpha_{\rm s}=1.5$) at position
$\boldsymbol{x}$ within our simulation volume and $f_{\heii{}}$ is the
\heii{} fraction in the \igm{} when \heii{} photoheating commences (we
assume $f_{\rm HeII}=1$).    
This expression assumes the excess energy
per \heii{} photoionization is shared with the baryons in the \igm{}
via Coulomb interactions.
Secondary ionizations by fast electrons
will only occur $\sim1$ per cent of the time and can be ignored here
\citep{Shull:1985p268,McQuinn:2009p7230}.

The last step is to then add the temperature boost from
equation~(\ref{eq:Tboost}) to an initial thermal state prior to \heii{}
reionization. We follow the procedure used in GBW11 and assume the
\igm{} temperature follows a well-defined power-law 
temperature--density relation at $1+\delta\leq 10$ (e.g. \citealt{HuiGnedin1997}),
and a constant temperature at higher overdensites where radiative
cooling becomes important, such that
\begin{eqnarray}  \label{eq:Tdens}
T_{\rm init} = T_{0} \begin{cases} (1+\delta_{\rm b})^{\gamma-1} & (1+\delta_{\rm b})\leq 10\\ 
  10^{\gamma-1} & (1+\delta_{\rm b})>10. \end{cases}
\end{eqnarray}
In this work we assume $T_{0} = 10\,000$~K, consistent with
measurements of the \igm{} thermal state at $z\sim 4$-$5$, prior to the
onset of \heii{} reionization
\citep{Schaye:2000p7569,Lidz:2010p7574,Becker:2011p7578}.  Note,
however, $\gamma$ is still poorly constrained at these redshifts; we
therefore adopt a value of $\gamma=1.3$ which lies in the middle of
the theoretically expected range of $\gamma=1.0$--$1.6$.

The \igm{} temperature in our patchy heating model is then computed by
adding the expected temperature boost from \heii{} photoheating to
equation~(\ref{eq:Tdens}),
\begin{eqnarray} \label{eq:TIGM_HeII}
T_{\igm{}} = T_{\rm init} + \Delta T(\boldsymbol{x}).
\end{eqnarray}
As we shall now demonstrate in the next section, this simple,
semi-analytical model allows us to qualitatively capture the effect of
large scale temperature fluctuations on the \lya{} forest without
resorting to expensive RT simulations.

\section{Thermal fluctuations in the \igm{} following \heii{} reionization}  \label{sec:IGMTemp}
\subsection{Predicted heating following \heii{} reionization}

\begin{figure*} 
	\begin{center}
		\includegraphics[trim = 0cm 7.5cm 0cm 7.5cm, scale = 0.85]{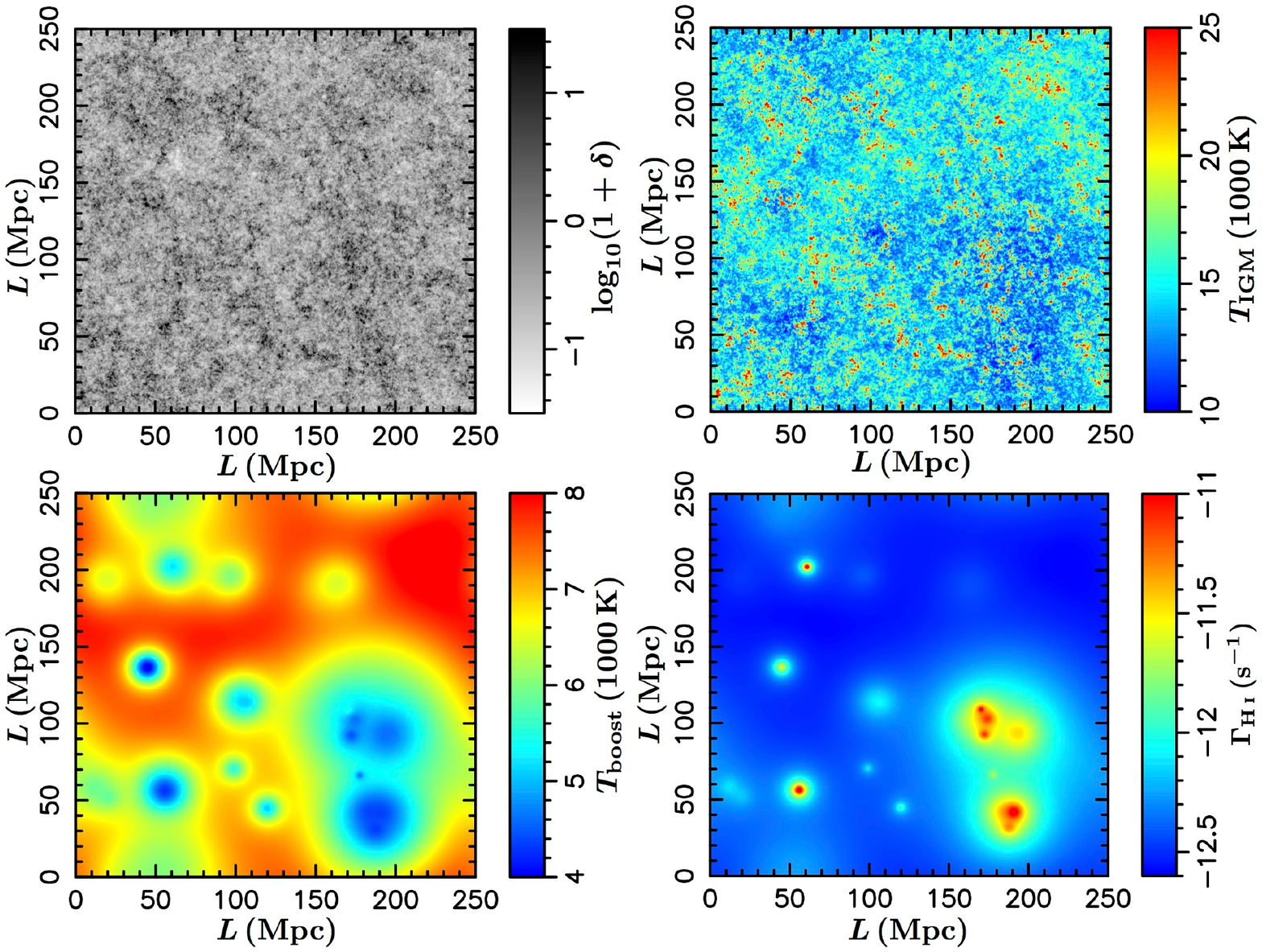}
	\end{center}
\caption[]{Slices through our \lya{} forest model at
  $z=2.5$. The slice thickness corresponds to a width of
    $\sim240$~kpc (i.e. our simulation resolution). Top
    left: the rank-ordered \igm{} density contrast (see GBW11 for
  details). Top right: the \igm{} temperature following
  \heii{} photoheating. Bottom left: the boost to the \igm{}
  temperature following \heii{} photoheating. Bottom right:
  the \hi{} photoionization rate.  The clustering of the ionizing
  sources (quasars) is clearly apparent in this panel.}
\label{fig:SlicePlots}
\end{figure*} 

Fig.\,\ref{fig:SlicePlots} displays the results of our simple model
for temperature fluctuations in $250\rm\,Mpc^{2}$ slices within our
simulation volume.  Clockwise from top left, the panels show the
rank-ordered non-linear \igm{} density (see GBW11 for details), the
\igm{} temperature following the addition of a temperature boost from
\heii{} photoheating, the \hi{} photoionization rate and the
magnitude of the temperature boost predicted by equation~(\ref{eq:Tboost}).

The first point to note is that the quasars -- which reside at the
positions where $\Gamma_{\hi{}}$ is largest in the lower-right panel
-- clearly trace the overdense regions observed in the upper-left
panel, as expected.  It is also evident that patchy \heii{}
photoheating modifies the \igm{} temperature substantially, with
fluctuations of the order of $\sim 10^{4}$~K on scales of several 
hundred~Mpc.  Regions in the direct vicinity of the quasars
are cooler, primarily because they are preferentially heated by softer
\heii{} ionizing photons with relatively short mean free paths.
Conversely, the furthest regions from the quasars are hottest as they
are preferentially ionized by high energy (hard) ionizing photons with
long mean free paths.  Qualitatively, this picture is similar to the
RT simulations of \citet{McQuinn:2009p7230} and
\citet{Compostella:2013p5745}, who also find regions furthest from the
quasars tend to be hottest following \heii{} reionization.  Note,
however, that in the RT simulations, the timing of \heii{}
reionization also plays a role -- regions close to quasars are ionized
first and have a longer time to cool adiabatically. This effect acts in addition to the
spectral filtering effect, and is not incorporated into the simple time-independent
implementation used here.

Quantitatively, we observe an average temperature boost of $\approx
4\,000$~K in the vicinity of the quasars.  Regions of the \igm{}
furthest from the quasars instead typically achieve $\Delta T \approx
8\,000 - 10\,000$~K.  For comparison, \citet{McQuinn:2009p7230} find
an average temperature boost at mean density of $\approx 12\,000$~K,
although some of the regions reionized last in their simulations
experience $\Delta T \gtrsim 30\,000$~K.  More recent simulations from
\citet{Compostella:2013p5745} find a slightly lower temperature boost
at mean density of $\approx 10\,000$~K.  Observational measurements of
the \igm{} temperature from the curvature of the transmitted flux in
the \lya{} forest also indicate a boost in the \igm{} temperature of
$\approx 8\,000$--$10\,000$~K (for an assumed $\gamma = 1.3$;
\citealt{Becker:2011p7578}).  Our simple model thus broadly reproduces
the expected level of heating in the \igm{} following \heii{}
reionization for gas at mean density.  Note, however, we do not
recover the $\Delta T \gtrsim 30\,000$~K heating observed in the
\igm{} in the simulations by \citet{McQuinn:2009p7230}, which is in
part due to the fact we have no sub-grid prescription for the
(uncertain) additional filtering of the ionizing radiation by dense
structures in the \igm{} (see e.g.\ \citealt{Meiksin:2012p13598}).  On
the other hand, as the volume filling factor of these very hot regions
remains low, and the largest temperature boosts are confined to dense
regions ($1+\delta>10$) which are not well probed by the \lya{} forest
at $z\sim 2.5$, this is unlikely to impact significantly on our
results.

\subsection{The temperature--density relation}

\begin{figure} 
	\begin{center}
	  \includegraphics[trim = 0cm 2cm 0cm 0cm, scale = 0.23]{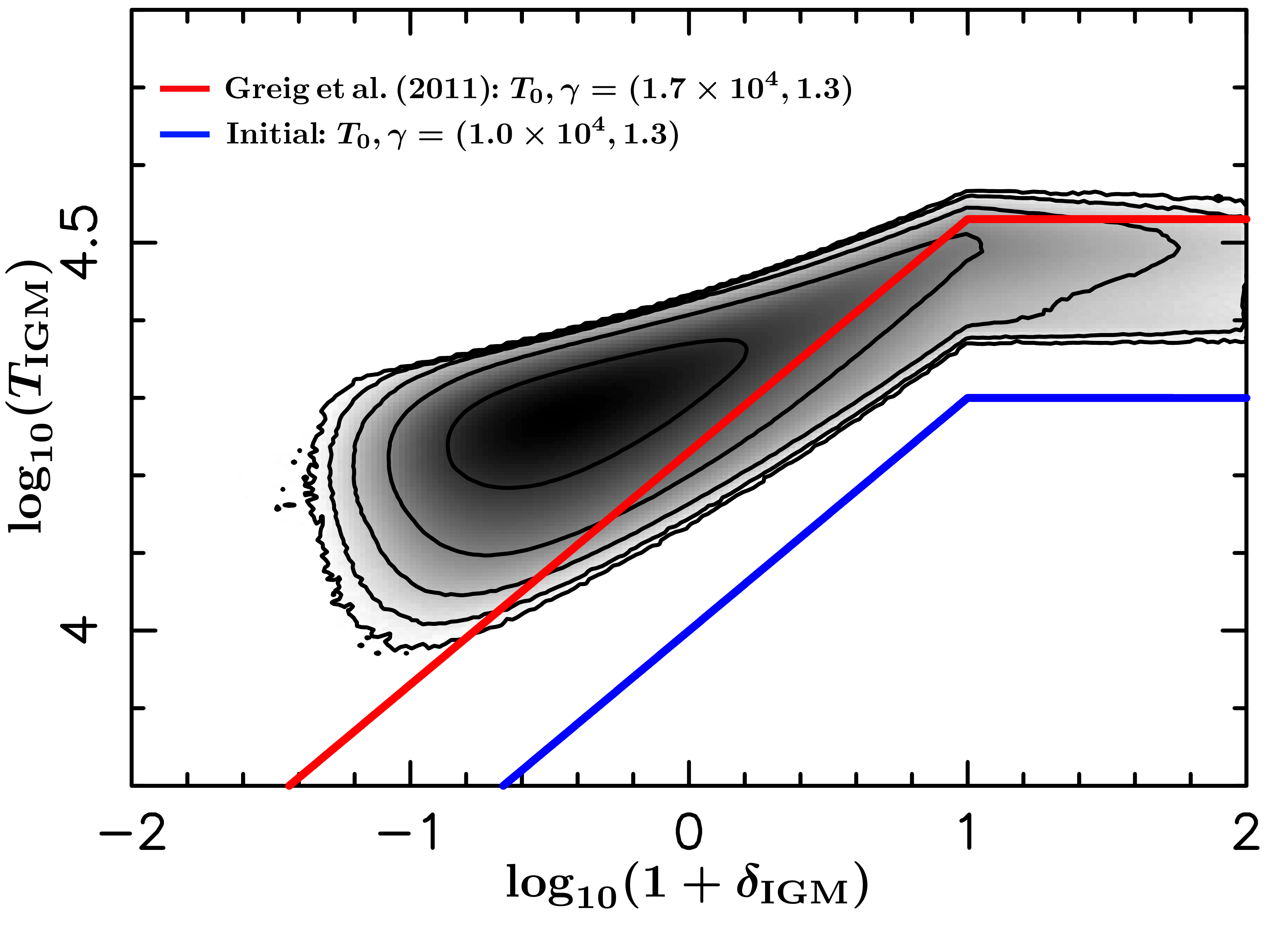}
        \end{center}
\caption[]{Contour plot of the temperature--density relation recovered
  from our \lya{} forest model at $z=2.5$.  The shaded region
  corresponds to the normalized frequency of the data while the solid
  contours represent 90, 70, 50, 30 and 10 per cent of this maximal
  value. The blue curve corresponds to $T_{\rm init}$, while the
  red curve corresponds to the temperature--density relation (without
  inhomogeneous \heii{} photoheating) used by GBW11.}
\label{fig:TdensScatter}
\end{figure}

RT simulations of \heii{} reionization also indicate
that soon after the commencement of \heii{} reionization, the \igm{}
temperature--density relation may take on a bimodal distribution
(\citealt{Bolton:2009p7229,McQuinn:2009p7230,Meiksin:2012p13598,Compostella:2013p5745}). The
recently photoheated gas will tend to follow a separate
temperature--density relation to the remainder of the \igm{} which is
yet to be reionized.  Following the completion of \heii{}
reionization, the bimodal \igm{} temperature--density relation
disappears and the gas starts to cool adiabatically.  However,
significant scatter (in excess of that expected from optically thin
models) and a shallower slope for the temperature--density relation
persist.

In Fig.\,~\ref{fig:TdensScatter}, we provide an example \igm{}
temperature--density scatter plot to demonstrate our simple model also
captures this effect following \heii{} reionization.  The results of
our simulation (contours and shaded region) are compared to the
assumed power-law expression adopted by GBW11 (red line). The boost in
the \igm{} temperature and the increased scatter in the relationship
following \heii{} reionization is clear on comparison to our initial
temperature--density relation, $T_{\rm init}$ (blue line).  There is
furthermore very good agreement between the temperature--density
relation in our semi-analytical model and the results from the
RT simulations of \citet{McQuinn:2009p7230} and
\citet{Compostella:2013p5745}.  We find, at all \igm{} densities, a
comparable level of scatter in the \igm{} temperature and similar
(i.e. shallower) slope for the temperature--density relation relative
to the value of $\gamma=1.3$ assumed prior to \heii{} photoheating.
Overall, these results give us confidence that our simple model will
capture the impact of \heii{} photoheating on the \igm{} adequately.

\section{The large-scale clustering of the \lya{} forest} \label{sec:Analysis}

In the previous sections, we have outlined our semi-analytical prescription for
incorporating large-scale temperature fluctuations into the fast,
efficient \lya{} forest simulations developed by GBW11.  We now shift
our focus to investigating the impact these fluctuations may have on
the \lya{} forest, paying particular attention to their effect on the
3D \lya{} forest power spectrum.

In what follows, we consider two separate cases. First, a model with
the \heii{} photoheating rates implemented as described above, and
secondly an otherwise identical model with the added \igm{} temperature
fluctuations using equation~(\ref{eq:TIGM_HeII}) removed.  In the
latter case, we assume a single \igm{} temperature--density relation,
with $T_{0} = 17\,000$~K and $\gamma = 1.3$.  In each case, we produce
10 independent realizations, enabling us to explore the box-to-box
scatter in our models (i.e. cosmic variance). The method for
producing \lya{} forest spectra from our simulations remains unchanged
from GBW11.
 In summary, we draw synthetic skewers of the 
\igm{} density, velocity and temperature fields from our semi-analytic model
and compute the total transmitted flux, $F=\mathrm{e}^{-\tau}$, where $\tau$ is the
\lya{} optical depth

\begin{equation}
\tau_{\rm{\alpha}}(i) =\frac{c\sigma_{\alpha}\delta R}{\upi^{1/2}}\sum_{j=1}^{N} \frac{n_{\hi{}}}{b_{\hi{}}(j)} \exp\left[-\left(\frac{v_{\h{}}(i)-u(j)}{b_{\hi{}}(j)}\right)^{2}\right].
\end{equation}
Here $i$ and $j$ denote pixels along the line of sight through the
simulation volume, $\delta R$ is the (proper) pixel width,
$\sigma_{\alpha} = 4.48\times10^{-18}\, \rm{cm}^{2}$ is the scattering
cross-section for $\lya$ photons, $b_{\hi{}} =
\left(\frac{2k_{\rm{B}}T}{m_{\rm{H}}}\right)^{1/2}$ is the Doppler
parameter describing the thermal width of the line profiles, $v_{\rm H}$
is the Hubble velocity and $u(j)$ is the total velocity given by the
summation of the Hubble flow and the peculiar velocity along the line
of sight, $u(j) = v_{\rm{H}}(j)+v_{\rm{pec}}(j)$. After constructing
our synthetic \lya{} forest spectra, we renormalize the pixel optical
depths to match the mean observed \lya{} flux at $z=2.5$ from the 
high-resolution sample of \citet{FaucherGiguere:2008p6637}.

\subsection{Synthetic \lya{} forest spectra}

\begin{figure*} 
	\begin{center}
		\includegraphics[trim = 0cm 0.8cm 0cm 0.3cm, scale = 0.75]{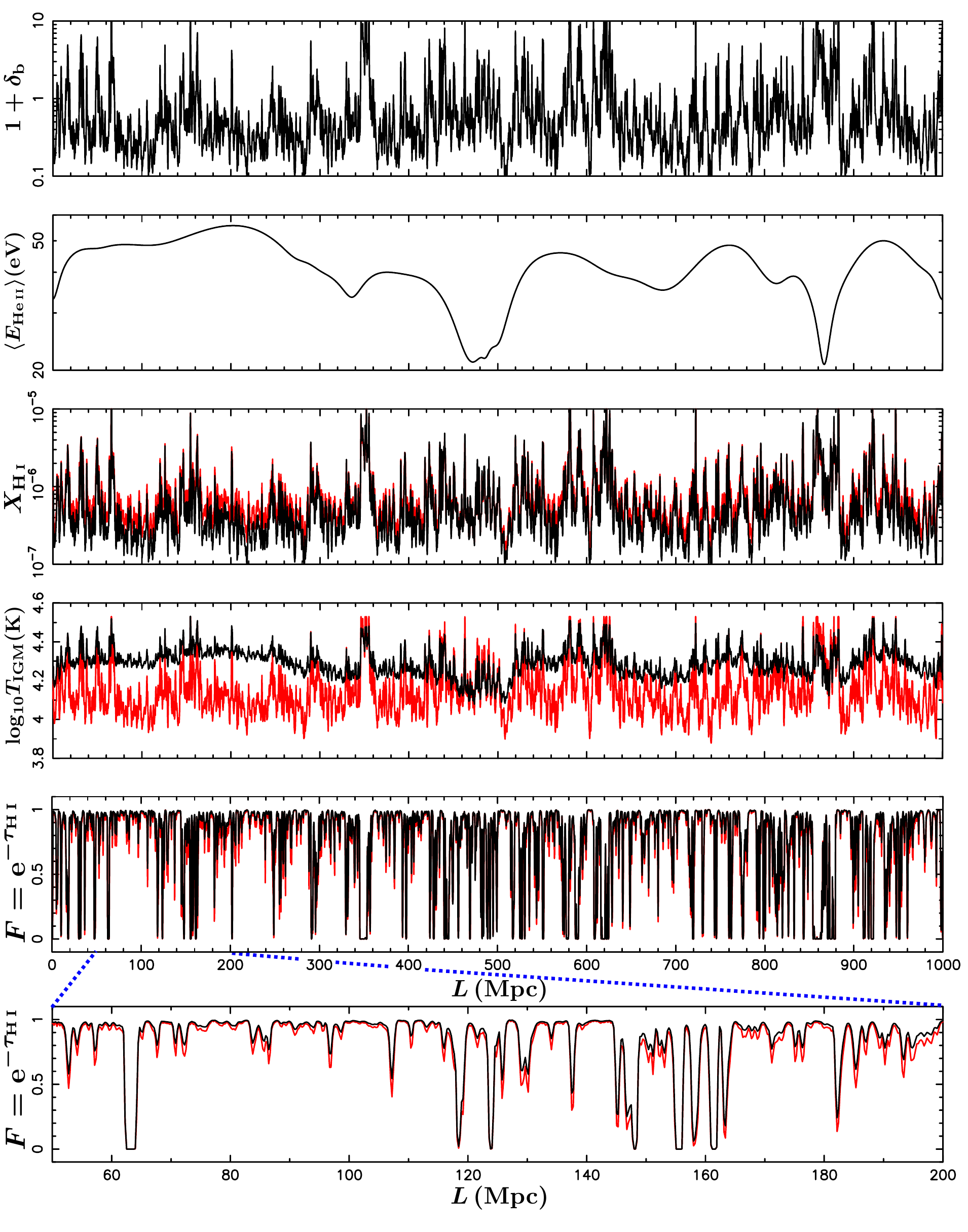}
	\end{center}
\caption[]{A line of sight drawn randomly through our \lya{} forest
  simulation.  From top to bottom, the panels display: the
  rank-ordered \igm{} density contrast; the average excess energy per
  \heii{} ionization; the \hi{} fraction $X_{\hi{}} = n_{\hi}/n_{\rm
    H}$; the gas temperature; the transmitted fraction $F=\mathrm{e}^{-\tau}$,
  and a 150~Mpc zoom in of the \lya{} forest spectrum in the previous
  panel. In all panels, the black (red) curves display models
  including (excluding) the large-scale temperature fluctuations
  computed using equation~(\ref{eq:Tboost}).}
\label{fig:LOS_figure}
\end{figure*}

In Fig.\,~\ref{fig:LOS_figure} we display a synthetic \lya{} forest
spectrum drawn through our 1\,Gpc$^{3}$ simulation volume.  In the
first panel, we show the line of sight \igm{} density contrast while
in the second we show the average excess energy per \heii{}
photoionization. This particular line of sight passes nearby several
quasars, with their locations corresponding to the minima where
$\langle E_{\rm HeII} \rangle\sim 20 \rm\, eV$.  These minima arise
due to heating close to the quasars being dominated by ionizing
photons near the \heii{} ionization edge which have shorter mean free
paths.  Additionally, $\langle E_{\rm HeII} \rangle\sim 50$--$60 \rm\,
eV$ furthest from the quasars, is only a factor of 2 lower
than the theoretical maximum value expected if every ionizing photon
were absorbed with equal probability, such that $\langle E\rangle
_{\rm thick}=h_{\rm p}\nu_{\rm HeII}/(\alpha_{\rm s}-1) = 108.8\rm\,eV$
(\citealt{Abel:1999pL13}).

In the third and fourth panels of Fig.\,\ref{fig:LOS_figure} we
compare the \hi{} fraction ($X_{\hi{}} = n_{\hi}/n_{\rm H}$) and \igm{}
temperature in our two scenarios. The red curves correspond to the
model without temperature fluctuations.  We observe that the \igm{}
temperature fluctuations have an important impact on the \hi{}
fraction through the temperature dependence of the \hii{}
recombination coefficient, $\alpha_{\hi{}}\propto T^{-0.7}$. The
largest difference between the models occurs in the regions where the
mean excess energy per photoionization is greatest, resulting in a
decrease in the \hi{} fraction by as much as 60 per cent.  As noted
earlier, the \igm{} temperature in our model with patchy \heii{}
photoheating is on average systematically higher compared to a single
temperature--density power-law relation.  Large-scale (greater than a
few hundred Mpc) variations result in the gas temperature, and are
driven by the heating from long range, hard ionizing photons.

The final two panels in Fig.\,\ref{fig:LOS_figure} provide a direct
comparison of the transmitted fraction in the \lya{} forest.  Note
that for the purpose of this comparison only we have not
normalized the spectra to the same effective optical depth.  As
expected there is less transmission overall from the model without
temperature fluctuations, although the differences are small and are
only apparent on large scales.  As noted by \cite{McQuinn:2011p10618},
it is for this reason that detecting $\Delta T \sim 10^{4}\rm\,K$
temperature fluctuations from \heii{} reionization is difficult using
1D line-of-sight \lya{} forest statistics based on a handful of quasar
spectra (see also
\citealt{Theuns2002,Lai:2006p14154,Lidz:2010p7574}). While we do
  not explicitly show here the impact of the \igm{} temperature
  fluctuations on the 1D \lya{} flux power spectrum and probability
  distribution function, we find that the level of variation between
  the models is small, and is at most of the order of 5 per cent.  Note,
  however, \citet{Lee:2011p1738} suggest an analysis of the 1D \lya{}
  transmission with threshold probability functions may yield
  additional sensitivity to these large \igm{} temperature
  fluctuations.

\subsection{The 3D \lya{} forest power spectrum}

\begin{figure*} 
	\begin{center}
		\includegraphics[trim = 0cm 0cm 0cm 0cm, scale = 0.5]{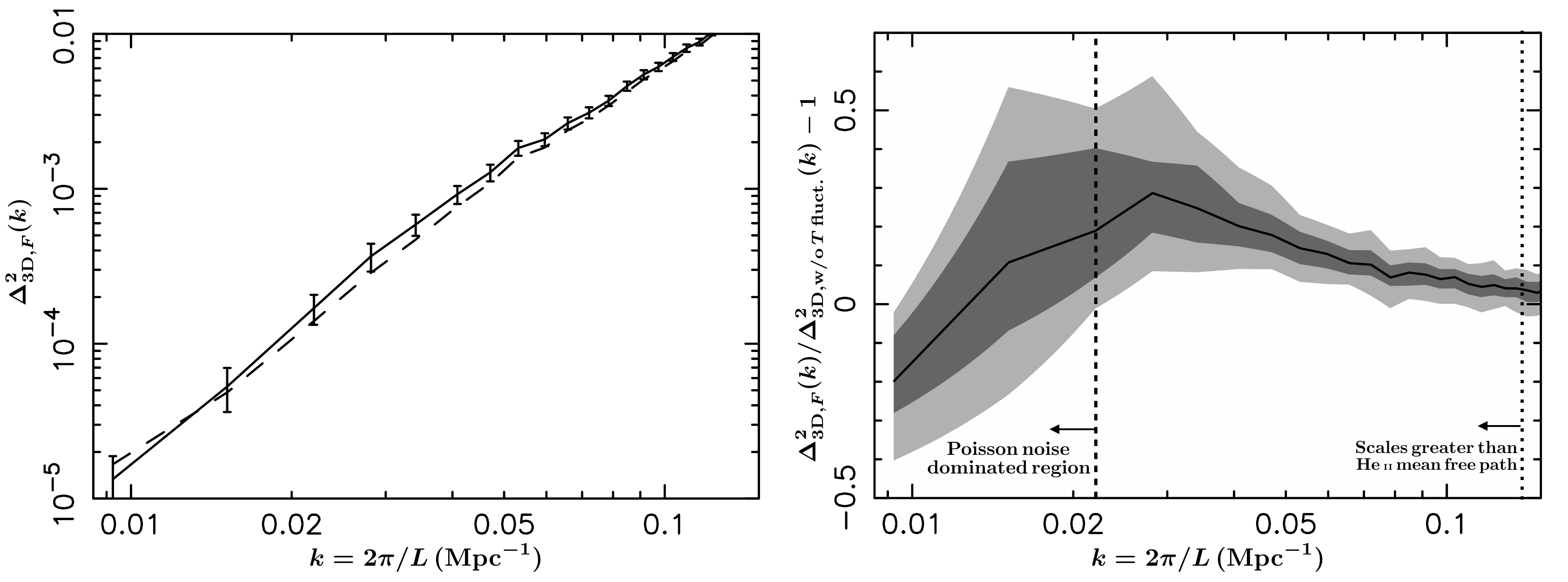}
	\end{center}
\caption[]{The reconstructed spherically averaged dimensionless 3D
  \lya{} forest power spectrum obtained from our simulations.
  Left-hand panel: the median power spectrum from our patchy
  \heii{} photoheating model (solid curve) and our model without
  \igm{} temperature fluctuations (dashed curve).  The error bars give
  the 1$\sigma$ Poisson errors on the number of Fourier modes within
  each spherical shell in $k$-space. Right-hand panel: the median
  fractional variation in the 3D power spectra obtained from our
  simulations. The shaded regions correspond to the predicted 68 and
  95 per cent scatter around the median.  Wavenumbers less than the
  vertical dashed line correspond to a Poisson noise dominated regime
  at scales approaching the box size, while wavenumbers left of the
  vertical dotted line denote scales greater than the mean free path
  at the \heii{} ionization edge ($\lambda_{\rm HeII} = 45$~Mpc)
  assumed in our model.}
\label{fig:Variation_Figure}

\end{figure*}

Large-scale temperature fluctuations are expected to have a small
effect on individual \lya{} forest spectra and on 1D line-of-sight
statistics.  However, in the case of the three-dimensional clustering
measurements from the \lya{} forest, temperature fluctuations may be
much more important (see also
\citealt{White:2010p5647,McQuinn:2011p10618}).  We thus estimate the
impact of the temperature fluctuations on the large-scale clustering
measurements of the \lya{} forest by taking the ratio of the
spherically averaged 3D \lya{} forest power spectrum from our
simulations with patchy \heii{} photoheating and a model without
temperature fluctuations.  Our reconstruction of the un-aliased
estimate of the 3D \lya{} forest spectrum (which arises due to the
sparsely sampled \lya{} forest number density) follows the method
outlined by \cite{McDonald2007} as applied in GBW11.

In this work, we assess our \lya{} forest modelling in the context of
the recent BOSS observations.  We therefore process our synthetic
spectra to mimic the BOSS instrumental resolution by convolving the
data with a Gaussian with an FWHM of $\sim 3.63\rm\, \AA$ followed by a
resampling on to bins of width $\sim 1.04\rm\,\AA$. We also use the
nominal mean survey design parameters of BOSS, where we measure our 3D
\lya{} forest power spectrum with a quasar density of $15\rm\,
deg^{-2}$ and add $\rm S/N = 5$ per pixel
\citep{Bolton:2012p7326,Dawson:2012p14037}.  Note also we only consider
line-of-sight sections of the forest corresponding to the distance
between the rest frame \lya{} and \lyb{} transitions at $z=2.5$
(roughly $\sim 620\rm\,Mpc$).  For each of the 10 realizations we
performed, we then generate $60\,000$ lines of sight within each
simulation box and randomly sample each at our prescribed
line-of-sight density to recover the 3D \lya{} forest power spectrum
100 times.  In this manner, we gain a good estimate of the impact of
large-scale temperature fluctuations within each simulation volume.

In the left-hand panel of Fig.\,\ref{fig:Variation_Figure} we show the median
spherically averaged 3D \lya{} forest power spectrum computed from our
two scenarios.  Here, the model with patchy \heii{} photoheating is
represented by the solid curve, while the model without temperature
fluctuations is given by the dashed curve.  Note again that to
construct these data, we recover the median power spectrum across all
of our 10 realizations. Therefore, these represent the median across a
sample of $1\,000$ individual power spectra.  An increase in power on
scales $k<0.1\rm\,Mpc^{-1}$ is evident in the model with patchy
heating.  The error bars show the 1$\sigma$ Poisson errors for a
single estimate of the spherically averaged 3D \lya{} forest power
spectrum.  At $k<0.02\rm\,Mpc^{-1}$, the limited number of $k$-space
bins sampled at these large scales results in significant Poisson
scatter which exceeds the difference between the two models (our box
size corresponds to $k=6.2\times 10^{-3}\rm\,Mpc^{-1}$, and the path
length between \lya{} and \lyb{} is $k\sim 0.01\rm\,Mpc^{-1}$).  Our
results for the 3D power spectrum at $k<0.02\rm\,Mpc^{-1}$ should
therefore be interpreted with caution.

Keeping this in mind, in the right panel of
Fig.\,\ref{fig:Variation_Figure} we show the fractional variation in
the median spherically averaged 3D \lya{} forest power spectrum relative to the
model excluding large-scale temperature fluctuations. The shaded
regions correspond to the 68 and 95 per cent scatter (dark and light
grey, respectively) around the median fractional variance. Furthermore,
the dotted line represents the wavenumber corresponding to the \heii{}
mean free path at $4\rm\,Ryd$ assumed in our model ($\lambda_{\heii{}}
= 45$~Mpc). Importantly, it is only on spatial scales significantly
larger than the mean free path at the ionization edge where the
temperature fluctuations have a notable impact on the large-scale
power.  This is in part because it is the long range heating by hard
photons which are responsible for producing the largest temperature
(and hence \hi{} fraction) fluctuations in our model.  At spatial
scales corresponding to $k\sim0.02$~Mpc$^{-1}$ the temperature
fluctuations increase large-scale power up to 20--30 per cent, with the
95 per cent scatter around the median showing an increase of up to
$\sim$60 per cent.  Furthermore, this is a scale-dependent effect.  At
decreasing spatial scales (larger $k$), the amplitude of the
temperature fluctuations is smaller, and the temperature fluctuations
have a limited effect on the power spectrum at $k>0.1\rm\,Mpc^{-1}$.
Note, however, that this inference is model dependent.  In particular,
if the mean free path for \heii{} ionizing photons is larger than we
have assumed, or if the sources responsible for reionizing \heii{} are
more numerous and/or less clustered, we expect the effect of
temperature fluctuations on the large-scale power to be smaller.  Our
results nevertheless suggest that with detailed forward modelling of
the 3D \lya{} forest power spectrum, it may be possible to probe the
inhomogeneous thermal state of the \igm{} following \heii{} reionization
with existing and forthcoming spectroscopic surveys.

In previous work, \citet{Pontzen:2014p0506} and
\citet{Gontcho:2014p7425} constructed analytical models that serve to
illustrate the dominant contributions to variations in the \hi{}
clustering power.  However, both studies focus on spatial fluctuations
in the ionizing background, which as discussed earlier will operate on
somewhat larger scales than the temperature fluctuations considered
here.  \citet{Gontcho:2014p7425} do briefly address the potential
effect of photoheating during \heii{} reionization, finding a rather
small impact on the \lya{} correlation function.  However, these
authors use the intensity of the \heii{} ionizing radiation (rather
than the mean excess energy per \heii{} ionization) to compute the
bias factor for thermal fluctuations in their model.  This may not
fully capture the important effect hard photons with long mean free
paths have on the photoheating rate.  On the other hand, using a
combination of numerical and analytical modelling
\citet{McQuinn:2011p10618} observe that temperature fluctuations
increase large-scale power by order unity on scales $k<
0.1\rm\,Mpc^{-1}$, again suggesting that our semi-analytical model
captures the effect of the temperature fluctuations reasonably well.

It is also worth noting that none of these models include the
  effect of winds on the IGM.  Recently \citet{Viel:2013p1743} have
  shown that AGN and supernova-driven winds can impact on 1D \lya{}
  forest flux statistics. While the magnitude of the effect is not
  large, the statistical power of the BOSS data will render these
  effects non-negligible.  How winds would impact upon the 3D \lya{}
  forest power spectrum on large scales, however, remains uncertain,
  although these effects are most prominent in the small-scale modes
  of the line-of-sight \lya{} forest power spectrum, and should occur
  on spatial scales smaller than the \igm{} temperature fluctuations
  considered here.

Finally, it is important to note that although we have only
  considered the 3D \lya{} forest power spectrum, astrophysical information
  may also be drawn from cross-correlations of the \lya{} forest with
  either DLAs \citep{FontRibera:2012p1742} or quasars
  \citep{FontRibera:2013p1739}.  Large-scale \igm{} temperature
  fluctuations could imprint a signature on to the quasar--\lya{} forest
  cross-correlation which is now measured to 80 $h^{-1}$~Mpc
  \citep{FontRibera:2013p1739}, although disentangling this signature
  from the quasar clustering and ionization fluctuations may be
  challenging.

\subsection{Recovery of the \bao{} scale}

We now briefly turn our attention to the impact of large-scale \igm{}
temperature fluctuations on the recovery of the \bao{} scale.  A
number of studies have already noted that fluctuations in the ionizing
background may add broad-band power to the 3D \lya{} forest correlation
function, which could be fitted and marginalized over by the addition of
a smoothed component
\citep{Slosar:2009p5710,McQuinn:2011p10618,Gontcho:2014p7425,Pontzen:2014p0506}. In
the case of strong ionizing background fluctuations both
\citet{Slosar:2009p5710} and \citet{Pontzen:2014p0506} observed that
these could shift the recovered \bao{} position by a few per cent,
although this is within the noise of the \bao{} measurements from BOSS
\citep{Busca:2012p13850,Slosar:2013p15454,Delubac:2014p1801,Ribera:2014p27}

Consequently, to similarly assess the impact of temperature
fluctuations on the recovery of the \bao{} scale, we closely follow the
approach used in GBW11.  We perform two concurrent \lya{} forest
simulations, one with and one without BAOs. We then take the ratio of
the 3D \lya{} forest power spectra predicted by these two simulations,
and recover an estimate of the \bao{} scale using a simple two
parameter model.  We perform this test for models including and
excluding patchy \heii{} photoheating.  In both cases, we confirm
there is little impact on the recovery of the \bao{} scale.  The 
large-scale temperature fluctuations result only in the addition of a smooth,
scale-dependent increase in power that can potentially be marginalized
over.  Nevertheless, as the \emph{significance} of the recovered BAO
peak is often inferred relative to the null hypothesis of no BAO peak,
this broad-band term is important.  Consequently, a complete forward
modelling of the broad-band term in the \lya{} correlation function
will require the incorporation of both ionizing background and
temperature fluctuations.

\section{Conclusions} \label{sec:conclusion}

The recovery of the large-scale clustering of the \igm{} and the
\bao{} feature from the \lya{} forest measured by BOSS
\citep{Busca:2012p13850,Slosar:2013p15454,Delubac:2014p1801,Ribera:2014p27}
provide a new avenue for precision cosmology from the \lya{} forest
at high-$z$.  In addition, the wealth of observational data also provides
a unique opportunity to explore the thermal state of the \igm{} following
the epoch of \heii{} reionization.  However, the impact of patchy
\heii{} photoheating on the \igm{} is subtle, requiring detailed
modelling to extract its signature from large spectroscopic data
sets. 

In this work, we have therefore developed a semi-analytical model for
investigating the impact of large-scale temperature fluctuations on
the three-dimensional clustering of the \lya{} forest. Our method is
built upon an existing efficient, large-volume, mock \lya{} forest
survey model developed by \cite{Greig:2011p13768}.  We have modified
this model to include several key ingredients required for modelling
the \igm{} following \heii{} reionization, including quasar clustering, a
spatially varying ionizing background and inhomogeneous \heii{}
photoheating.

We find our simple prescription for patchy \heii{} photoheating is in
very good agreement with the results from recent RT
simulations of \heii{} reionization
(\citealt{McQuinn:2009p7230,Compostella:2013p5745}). However, the
dynamic range of our modelling is substantially larger than currently
possible with these more expensive, fully numerical approaches.  
Large-scale temperature fluctuations typically produce a 20--30 per cent
increase in the 3D spherically averaged \lya{} forest power spectrum at
$k\sim 0.02\rm\,Mpc^{-1}$, although scatter arising from cosmic
variance indicates this can be as large as 50--60 per cent.
Importantly, we also observe this to be a scale-dependent effect with
an amplitude that increases towards larger spatial scales.
Furthermore, this effect is expected to occur at scales larger than
both the mean quasar separation and the mean free path of \heii{}
ionizing photons at the photoelectric edge.  We note, however, that
our results are model dependent.  An increase in the mean free path
for \heii{} ionizing photons, or more numerous or less clustered
ionizing sources will result in a reduction in the increase of 
large-scale power. Nevertheless, with detailed forward modelling, it may be
possible to probe the inhomogeneous thermal state of the \igm{} following
\heii{} reionization with existing and forthcoming spectroscopic
surveys.

Finally, we briefly consider the potential systematic uncertainty that
large-scale temperature fluctuations may have on the recovery of the
\bao{} scale. We find that these do not substantially impact upon the
recovery of the \bao{} scale; large-scale temperature fluctuations
result only in the addition of smooth, scale-dependent increase in
power that can potentially be marginalized over.  However, any
complete forward modelling of the broad-band term in the \lya{}
correlation function will ultimately require the incorporation of both
ionizing background
\citep{Slosar:2009p5710,White:2010p5647,Gontcho:2014p7425,Pontzen:2014p0506}
\emph{and} large-scale thermal fluctuations.

\section*{Acknowledgements}

We thank Andrew Pontzen, Satya Gontcho A Gontcho and K.G.\ Lee
for providing helpful comments on a draft version of this manuscript. We
also thank the anonymous referee for providing helpful suggestions.
BG acknowledges the support of the Australian Postgraduate Award.  The
Centre for All-sky Astrophysics is an Australian Research Council
Centre of Excellence, funded by grant CE110001020.  JSB acknowledges
the support of a Royal Society University Research Fellowship.

\bibliography{Papers.bib}

\begin{thebibliography}{63}
\expandafter\ifx\csname natexlab\endcsname\relax\def\natexlab#1{#1}\fi

\bibitem[{Abel \& Haehnelt(1999)}]{Abel:1999pL13}
Abel T., Haehnelt M.~G., 1999, ApJ, 520, L13

\bibitem[{Becker {et~al.}(2011)Becker, Bolton, Haehnelt, \&
  Sargent}]{Becker:2011p7578}
Becker G.~D., Bolton J.~S., Haehnelt M.~G., Sargent W. L.~W., 2011, MNRAS, 410,
  1096

\bibitem[{Bolton {et~al.}(2012)}]{Bolton:2012p7326}
Bolton A.~S., {et~al.}, 2012, AJ, 144, 144

\bibitem[{Bolton {et~al.}(2010)Bolton, Becker, Wyithe, Haehnelt, \&
  Sargent}]{Bolton:2010p6750}
Bolton J.~S., Becker G.~D., Wyithe J. S.~B., Haehnelt M.~G., Sargent W. L.~W.,
  2010, MNRAS, 406, 612

\bibitem[{Bolton {et~al.}(2009)Bolton, Oh, \& Furlanetto}]{Bolton:2009p7229}
Bolton J.~S., Oh S.~P., Furlanetto S.~R., 2009, MNRAS, 395, 736

\bibitem[{Bolton \& Viel(2011)}]{Bolton:2011p13588}
Bolton J.~S., Viel M., 2011, MNRAS, 414, 241

\bibitem[{Busca {et~al.}(2013)}]{Busca:2012p13850}
Busca N.~G., {et~al.}, 2013, A\&A, 552, 96

\bibitem[{Compostella {et~al.}(2013)Compostella, Cantalupo, \&
  Porciani}]{Compostella:2013p5745}
Compostella M., Cantalupo S., Porciani C., 2013, MNRAS, 435, 3169

\bibitem[{Croft {et~al.}(2002)Croft, Weinberg, Bolte, Burles, Hernquist, Katz,
  Kirkman, \& Tytler}]{Croft:2002p6751}
Croft R. A.~C., Weinberg D.~H., Bolte M., Burles S., Hernquist L., Katz N.,
  Kirkman D., Tytler D., 2002, ApJ, 581, 20

\bibitem[{Croft {et~al.}(1998)Croft, Weinberg, Katz, \&
  Hernquist}]{Croft:1998p44}
Croft R. A.~C., Weinberg D.~H., Katz N., Hernquist L., 1998, ApJ, 495, 44

\bibitem[{Davies \& Furlanetto(2014)}]{Davies:2012p14012}
Davies F.~B., Furlanetto S.~R., 2014, MNRAS, 437, 1141

\bibitem[{Dawson {et~al.}(2013)}]{Dawson:2012p14037}
Dawson K.~S., {et~al.}, 2013, AJ, 145, 10

\bibitem[{Delubac {et~al.}(2014)}]{Delubac:2014p1801}
Delubac T., {et~al.}, 2014, preprint (arXiv:1404.1801)

\bibitem[{Fardal {et~al.}(1998)Fardal, Giroux, \& Shull}]{Fardal:1998p2206}
Fardal M.~A., Giroux M.~L., Shull J.~M., 1998, AJ, 115, 2206

\bibitem[{Faucher-Gigu{\`e}re {et~al.}(2008)Faucher-Gigu{\`e}re, Prochaska,
  Lidz, Hernquist, \& Zaldarriaga}]{FaucherGiguere:2008p6637}
Faucher-Gigu{\`e}re C.-A., Prochaska J.~X., Lidz A., Hernquist L., Zaldarriaga
  M., 2008, ApJ, 681, 831

\bibitem[{Font-Ribera {et~al.}(2012{\natexlab{a}})Font-Ribera, McDonald, \&
  Miralda-Escud\'{e}}]{FontRibera:2012p1}
Font-Ribera A., McDonald P., Miralda-Escud\'{e} J., 2012{\natexlab{a}}, J.
  Cosmol. Astropart. Phys., 01, 001

\bibitem[{Font-Ribera {et~al.}(2012{\natexlab{b}})}]{FontRibera:2012p1742}
Font-Ribera A., {et~al.}, 2012{\natexlab{b}}, J. Cosmol. Astropart. Phys., 11,
  059

\bibitem[{Font-Ribera {et~al.}(2013)}]{FontRibera:2013p1739}
---, 2013, J. Cosmol. Astropart. Phys., 05, 018

\bibitem[{Font-Ribera {et~al.}(2014)}]{Ribera:2014p27}
---, 2014, J. Cosmol. Astropart. Phys., 5, 27

\bibitem[{Furlanetto(2009)}]{Furlanetto:2009p13603}
Furlanetto S.~R., 2009, ApJ, 703, 702

\bibitem[{Furlanetto \& Oh(2008{\natexlab{a}})}]{Furlanetto:2008p4660}
Furlanetto S.~R., Oh S.~P., 2008{\natexlab{a}}, ApJ, 681, 1

\bibitem[{Furlanetto \& Oh(2008{\natexlab{b}})}]{Furlanetto:2008p13605}
---, 2008{\natexlab{b}}, ApJ, 682, 14

\bibitem[{{Gontcho A Gontcho} {et~al.}(2014){Gontcho A Gontcho},
  Miralda-Escud\'{e}, \& Busca}]{Gontcho:2014p7425}
{Gontcho A Gontcho} S., Miralda-Escud\'{e} J., Busca N.~G., 2014, MNRAS, 442,
  187

\bibitem[{Greig {et~al.}(2011)Greig, Bolton, \& Wyithe}]{Greig:2011p13768}
Greig B., Bolton J.~S., Wyithe J. S.~B., 2011, MNRAS, 418, 1980, (GBW11)

\bibitem[{Haardt \& Madau(2012)}]{HaardtMadau2012}
Haardt F., Madau P., 2012, ApJ, 746, 125

\bibitem[{Hopkins {et~al.}(2007)Hopkins, Richards, \&
  Hernquist}]{Hopkins:2007p13715}
Hopkins P.~F., Richards G.~T., Hernquist L., 2007, ApJ, 654, 731

\bibitem[{Hui \& Gnedin(1997)}]{HuiGnedin1997}
Hui L., Gnedin N.~Y., 1997, MNRAS, 292, 27

\bibitem[{Khaire \& Srianand(2013)}]{Khaire2013}
Khaire V., Srianand R., 2013, MNRAS, 431, L53

\bibitem[{Lai {et~al.}(2006)Lai, Lidz, Hernquist, \&
  Zaldarriaga}]{Lai:2006p14154}
Lai K., Lidz A., Hernquist L., Zaldarriaga M., 2006, ApJ, 644, 61

\bibitem[{{Le Goff} {et~al.}(2011)}]{LeGoff:2011p1737}
{Le Goff} J.~M., {et~al.}, 2011, A\&A, 534, A135

\bibitem[{Lee \& Spergel(2011)}]{Lee:2011p1738}
Lee K.-G., Spergel D.~N., 2011, ApJ, 734, 21

\bibitem[{Lidz {et~al.}(2010)Lidz, Faucher-Gigu{\`e}re, Dall'Aglio, McQuinn,
  Fechner, Zaldarriaga, Hernquist, \& Dutta}]{Lidz:2010p7574}
Lidz A., Faucher-Gigu{\`e}re C.-A., Dall'Aglio A., McQuinn M., Fechner C.,
  Zaldarriaga M., Hernquist L., Dutta S., 2010, ApJ, 718, 199

\bibitem[{Lidz \& Malloy(2014)}]{LidzMalloy2014}
Lidz A., Malloy M., 2014, ApJ, 788, 175

\bibitem[{{Luki{\'c}} {et~al.}(2015){Luki{\'c}}, {Stark}, {Nugent}, {White},
  {Meiksin}, \& {Almgren}}]{Lukic2014}
{Luki{\'c}} Z., {Stark} C., {Nugent} P., {White} M., {Meiksin} A., {Almgren}
  A., 2015, MNRAS, 446, 3697

\bibitem[{Madau {et~al.}(1999)Madau, Haardt, \& Rees}]{Madau:1999p13717}
Madau P., Haardt F., Rees M.~J., 1999, ApJ, 514, 648

\bibitem[{McDonald \& Eisenstein(2007)}]{McDonald2007}
McDonald P., Eisenstein D.~J., 2007, Phys. Rev. D, 76, 063009

\bibitem[{McDonald {et~al.}(2006)}]{McDonald:2006p6764}
McDonald P., {et~al.}, 2006, ApJS, 163, 80

\bibitem[{McQuinn {et~al.}(2011)McQuinn, Hernquist, Lidz, \&
  Zaldarriaga}]{McQuinn:2011p10618}
McQuinn M., Hernquist L., Lidz A., Zaldarriaga M., 2011, MNRAS, 415, 977

\bibitem[{McQuinn {et~al.}(2009)McQuinn, Lidz, Zaldarriaga, Hernquist, Hopkins,
  Dutta, \& Faucher-Gigu{\`e}re}]{McQuinn:2009p7230}
McQuinn M., Lidz A., Zaldarriaga M., Hernquist L., Hopkins P.~F., Dutta S.,
  Faucher-Gigu{\`e}re C.-A., 2009, ApJ, 694, 842

\bibitem[{Meiksin \& Tittley(2012)}]{Meiksin:2012p13598}
Meiksin A., Tittley E.~R., 2012, MNRAS, 423, 7

\bibitem[{O'Meara {et~al.}(2013)O'Meara, Prochaska, Worseck, Chen, \&
  Madau}]{OMeara:2013p137}
O'Meara J.~M., Prochaska J.~X., Worseck G., Chen H.-W., Madau P., 2013, ApJ,
  765, 137

\bibitem[{Palanque-Delabrouille {et~al.}(2013)}]{Palanque-Delabrouille2013}
Palanque-Delabrouille N., {et~al.}, 2013, A\&A, 559, A85

\bibitem[{Peirani {et~al.}(2014)Peirani, Weinberg, Colombi, Blaizot, Dubois, \&
  Pichon}]{Peirani:2014p1703}
Peirani S., Weinberg D.~H., Colombi S., Blaizot J., Dubois Y., Pichon C., 2014,
  ApJ, 784, 11

\bibitem[{{Planck Collaboration XVI}(2014)}]{PlanckCollaboration:2013p15401}
{Planck Collaboration XVI}, 2014, A\&A, 571, A16

\bibitem[{Pontzen(2014)}]{Pontzen:2014p0506}
Pontzen A., 2014, Phys. Rev. D, 89, 083010

\bibitem[{Raskutti {et~al.}(2012)Raskutti, Bolton, Wyithe, \&
  Becker}]{Raskutti2012}
Raskutti S., Bolton J.~S., Wyithe J. S.~B., Becker G.~D., 2012, MNRAS, 421,
  1969

\bibitem[{Rauch(1998)}]{Rauch:1998p4563}
Rauch M., 1998, ARA\&A, 36, 267

\bibitem[{Rudie {et~al.}(2013)Rudie, Steidel, Shapley, \& Pettini}]{Rudie2013}
Rudie G.~C., Steidel C.~C., Shapley A.~E., Pettini M., 2013, ApJ, 769, 146

\bibitem[{Schaye {et~al.}(2000)Schaye, Theuns, Rauch, Efstathiou, \&
  Sargent}]{Schaye:2000p7569}
Schaye J., Theuns T., Rauch M., Efstathiou G., Sargent W. L.~W., 2000, MNRAS,
  318, 817

\bibitem[{Seljak {et~al.}(2006)Seljak, Slosar, \& McDonald}]{Seljak:2006p14}
Seljak U., Slosar A., McDonald P., 2006, J. Cosmol. Astropart. Phys., 10, 014

\bibitem[{Shen {et~al.}(2007)}]{Shen:2007p2222}
Shen Y., {et~al.}, 2007, AJ, 133, 2222

\bibitem[{Shull {et~al.}(2012)Shull, Stevans, \& Danforth}]{Shull:2012p162}
Shull J.~M., Stevans M., Danforth C.~W., 2012, ApJ, 752, 162

\bibitem[{Shull \& {van Steenberg}(1985)}]{Shull:1985p268}
Shull J.~M., {van Steenberg} M.~E., 1985, ApJ, 298, 268

\bibitem[{Slosar {et~al.}(2009)Slosar, Ho, White, \& Louis}]{Slosar:2009p5710}
Slosar A., Ho S., White M., Louis T., 2009, J. Cosmol. Astropart. Phys., 10,
  019

\bibitem[{Slosar {et~al.}(2013)}]{Slosar:2013p15454}
Slosar A., {et~al.}, 2013, J. Cosmol. Astropart. Phys., 04, 026

\bibitem[{Stevans {et~al.}(2014)Stevans, Shull, Danforth, \&
  Tilton}]{Stevans2014}
Stevans M.~L., Shull J.~M., Danforth C.~W., Tilton E.~M., 2014, ApJ, 794, 75

\bibitem[{Telfer {et~al.}(2002)Telfer, Zheng, Kriss, \&
  Davidsen}]{Telfer:2002p14077}
Telfer R.~C., Zheng W., Kriss G.~A., Davidsen A.~F., 2002, ApJ, 565, 773

\bibitem[{Theuns {et~al.}(1998)Theuns, Leonard, Efstathiou, Pearce, \&
  Thomas}]{Theuns:1998p7245}
Theuns T., Leonard A., Efstathiou G., Pearce F.~R., Thomas P.~A., 1998, MNRAS,
  301, 478

\bibitem[{Theuns {et~al.}(2002)Theuns, Zaroubi, Kim, Tzanavaris, \&
  Carswell}]{Theuns2002}
Theuns T., Zaroubi S., Kim T.-S., Tzanavaris P., Carswell R.~F., 2002, MNRAS,
  332, 367

\bibitem[{Viel {et~al.}(2004)Viel, Haehnelt, \& Springel}]{Viel:2004p7045}
Viel M., Haehnelt M.~G., Springel V., 2004, MNRAS, 354, 684

\bibitem[{Viel {et~al.}(2002)Viel, Matarrese, Mo, Theuns, \&
  Haehnelt}]{Viel:2002p4288}
Viel M., Matarrese S., Mo H.~J., Theuns T., Haehnelt M.~G., 2002, MNRAS, 336,
  685

\bibitem[{Viel {et~al.}(2013)Viel, Schaye, \& Booth}]{Viel:2013p1743}
Viel M., Schaye J., Booth C.~M., 2013, MNRAS, 429, 1734

\bibitem[{White {et~al.}(2010)White, Pope, Carlson, Heitmann, Habib, Fasel,
  Daniel, \& Lukic}]{White:2010p5647}
White M., Pope A., Carlson J., Heitmann K., Habib S., Fasel P., Daniel D.,
  Lukic Z., 2010, ApJ, 713, 383

\end{thebibliography}

\end{document}